\documentclass[conference]{IEEEtran}
\IEEEoverridecommandlockouts

\usepackage[T1]{fontenc}
\usepackage{textcomp}
\usepackage{blindtext}
\usepackage[utf8]{inputenc}
\usepackage[noadjust]{cite}
\usepackage[font=footnotesize]{caption}
\usepackage{dblfloatfix}
\usepackage{graphicx}
\usepackage{psfrag}
\usepackage{subfig}
\usepackage{amsmath,amssymb,amsthm,mathrsfs,amsfonts,dsfont}
\usepackage{color}
\usepackage[algoruled]{algorithm2e}
\usepackage{fixltx2e}
\usepackage{multirow}
\usepackage{multicol}
\usepackage[hidelinks]{hyperref}
\usepackage[normalem]{ulem}
\usepackage{acronym}
\usepackage[table,xcdraw]{xcolor}
\usepackage{accents}
\usepackage{mwe}
\usepackage{hhline}
\usepackage{trfsigns}
\usepackage{siunitx}
\usepackage{lscape}
\usepackage{relsize}

\definecolor{colorSC}{RGB}{158,100,120}

\newacro{5G}{fifth generation}
\newacro{6G}{sixth generation}
\newacro{A/D}{analog-to-digital}
\newacro{ADC}{analog-to-digital converter}
\newacro{AFE}{analog front-end}
\newacro{AWGN}{additive white Gaussian noise}
\newacro{B5G}{beyond \ac{5G}}
\newacro{BER}{bit error ratio}
\newacro{BPSK}{binary phase-shift keying}
\newacro{BP}{band-pass}
\newacro{CCDF}{complementary cumulative distribution function}
\newacro{CDM}{code-division multiplexing}
\newacro{CFO}{carrier frequency offset}
\newacro{CFR}{channel frequency response}
\newacro{CIR}{channel impulse response}
\newacro{CP}{cyclic prefix}
\newacro{CPO}{carrier phase offset}
\newacro{CR}[C\&R]{wireless communication and radar sensing}
\newacro{CS}{chirp sequence}
\newacro{CSI}{channel state information}
\newacro{CW}{continuous wave}
\newacro{D/A}{digital-to-analog}
\newacro{DAC}{digital-to-analog converter}
\newacro{DDS}{direct digital synthesis}
\newacro{DFCS}{dual-functional communication and radar sensing}
\newacro{DFRC}{dual-functional radar-communication}
\newacro{DFnT}{discrete Fresnel transform}
\newacro{DFT}{discrete Fourier transform}
\newacro{DMRS}{demodulation reference signal}
\newacro{DoA}{direction of arrival}
\newacro{ETI}{Emerging Technology Initiative}
\newacro{ETSI}{European Telecommunications Standards Institute}
\newacro{EVM}{error vector magnitude}
\newacro{FBMC}{filter bank multicarrier}
\newacro{FCC}{Federal Communications Commission}
\newacro{FDE}{frequency-domain equalization}
\newacro{FDM}{frequency-division multiplexing}
\newacro{FDZP}{frequency-domain zero padding}
\newacro{FIR}{finite impulse response}
\newacro{FMCW}{frequency-modulated continuous wave}
\newacro{FPGA}{field programmable gate array}
\newacro{FrDM}{Fresnel-division multiplexing}
\newacro{FSK}{frequency-shift keying}
\newacro{FSP}{frequency-shift precoding}
\newacro{GFDM}{generalized frequency-division multiplexing}
\newacro{HAD}{highly automated driving}
\newacro{HP}{high-pass}
\newacro{IBFD}{in-band full duplex}
\newacro{IC}{interference cancellation}
\newacro{ICI}{intercarrier interference}
\newacro{IDFT}{inverse discrete Fourier transform}
\newacro{IDFnT}{inverse discrete Fresnel transform}
\newacro{IEEE}{Institute of Electrical and Electronics Engineers}
\newacro{IF}{intermediate frequency}
\newacro{IM}{index modulation}
\newacro{ISAC}{integrated sensing and communication}
\newacro{ISI}{intersymbol interference}
\newacro{ISLR}{integrated-sidelobe level ratio}
\newacro{I/Q}{in-phase/quadrature}
\newacro{JCAS}{joint communication and sensing}
\newacro{JRC}{joint radar-communications}
\newacro{LDPC}{low-density parity-check}
\newacro{LNA}{low-noise amplifier}
\newacro{LO}{local oscillator}
\newacro{LoS}{line-of-sight}
\newacro{LFSR}{linear-feedback shift register}
\newacro{LP}{low-pass}
\newacro{LRR}{long range radar}
\newacro{mmWave}{milimeter wave}
\newacro{MER}{modulation error ratio}
\newacro{MIMO}{multiple-input multiple-output}
\newacro{MLS}{maximum-length sequence}
\newacro{NBI}{narrowband interference}
\newacro{NLoS}{non-line-of-sight}
\newacro{NR}{New Radio}
\newacro{OCDM}{orthogonal chirp-division multiplexing}
\newacro{OFDM}{orthogonal frequency-division multiplexing}
\newacro{OOB}{out-of-band}
\newacro{OTFS}{orthogonal time-frequency space}
\newacro{P/S}{paralell-to-serial}
\newacro{PA}{power amplifier}
\newacro{PACF}{periodic autocorrelation function}
\newacro{PAPR}{peak-to-average power ratio}
\newacro{PCCF}{periodic cross-correlation function}
\newacro{PL}{payload}
\newacro{PLC}{powerline communication}
\newacro{PLL}{phase-locked loop}
\newacro{PMCW}{phase-modulated continuous wave}
\newacro{PMEPR}{peak-to-mean envelope power ratio}
\newacro{PPLR}{peak power loss ratio}
\newacro{PRBS}{pseudorandom binary sequence}
\newacro{PSLR}{peak-to-sidelobe level ratio}
\newacro{QPSK}{quadrature phase-shift keying}
\newacro{RaaS}{radar as a service}
\newacro{RAT}{radio access technology}
\newacro{RadCom}{radar-communication}
\newacro{RCS}{radar cross section}
\newacro{RF}{radio-frequency}
\newacro{RTS}{radar target simulator}
\newacro{S/P}{serial-to-paralell}
\newacro{SC}[S\&C]{Schmidl \& Cox}
\newacro{SCO}{sampling clock offset}
\newacro{SDM}{spatial division multiplexing}
\newacro{SDMA}{spatial division multiple access}
\newacro{SFO}{sampling frequency offset}
\newacro{SH}[S\&H]{sample and hold}
\newacro{SI}{self-interference}
\newacro{SIC}{self-interference cancellation}
\newacro{SIR}{signal-to-interference ratio}
\newacro{SISO}{single-input single-output}
\newacro{SNR}{signal-to-noise ratio}
\newacro{SRR}{short range radar}
\newacro{SoC}{system-on-a-chip}
\newacro{STO}{symbol time offset}
\newacro{TDD}{time-division duplexing}
\newacro{TDE}{time-domain equalization}
\newacro{TDL}{tapped delay line}
\newacro{TDM}{time-division multiplexing}
\newacro{TDR}{time-domain reflectometry}
\newacro{UAV}{unmanned aerial vehicle}
\newacro{UE}{user equipment}
\newacro{UWAC}{underwater acoustic communication}
\newacro{V2I}{vehicle-to-infrastructure}
\newacro{V2V}{vehicle-to-vehicle}
\newacro{V2X}{vehicle-to-everything}
\newacro{ZF}{zero forcing}

\makeatletter
\renewcommand*\env@cases[1][1.2]{%
	\let\@ifnextchar\new@ifnextchar
	\left\lbrace
	\def\arraystretch{#1}%
	\array{@{}l@{\quad}l@{}}%
}
\makeatother

\usepackage{etoolbox}
\makeatletter
\patchcmd{\@makecaption}
{\scshape}
{}
{}
{}
\makeatother

\def\BibTeX{{\rm B\kern-.05em{\sc i\kern-.025em b}\kern-.08em
    T\kern-.1667em\lower.7ex\hbox{E}\kern-.125emX}}

\usepackage{fancyhdr}
\fancypagestyle{empty}{fancy}

\begin{document}

\title{
Enabling Joint Radar-Communication Operation in Shift Register-Based PMCW Radars
}

\author{\IEEEauthorblockN{Lucas Giroto de Oliveira$^*$, Elizabeth Bekker, Axel Diewald, Benjamin Nuss,\\ Theresa Antes, Yueheng Li, Akanksha Bhutani, and Thomas Zwick}
\IEEEauthorblockA{Institute of Radio Frequency Engineering and Electronics (IHE)\\
Karlsruhe Institute of Technology (KIT), Germany \\
E-mail: $^*$lucas.oliveira@kit.edu
}
}

\maketitle

\begin{abstract}
	This article introduces adaptations to the conventional frame structure in binary phase-modulated continuous wave (PMCW) radars with sequence generation via linear-feedbck shift registers and additional processing steps to enable joint radar-communication (RadCom) operation. In this context, a preamble structure based on pseudorandom binary sequences (PRBSs) that is compatible with existing synchronization algorithms is outlined, and the allocation of pilot PRBS blocks is discussed. Finally, results from proof-of-concept measurements are presented to illustrate the effects of the choice of system and signal parameters and validate the investigated PMCW-based RadCom system and synchronization strategy.
\end{abstract}


\section{Introduction}\label{sec:introduction}

Due to its low linearity requirements and high performance as a modulation scheme for fully digital radar systems, binary \ac{PMCW} has continuously gained interest for \ac{HAD} applications. In this context, using \acp{LFSR} instead of \acp{DAC} to generate transmit \acp{PRBS} appears as an efficient approach to achieve a simplified transmitter architecture while keeping a robust radar performance \cite{giroto2022_PMCW_MDPI}. With minor adaptations to the \ac{PMCW} transmitter \cite{probst2022}, joint \ac{RadCom} operation with the transmission of \ac{BPSK} symbols as described in \cite{giroto2021_tmtt} is also possible. This results in high communication robustness due to the experienced processing gain from accumulation and correlation, but in rather low data rates. 
The latter are, however, sufficient for radar-centric operation of the \ac{PMCW}-based \ac{RadCom} system, in which basic control information is exchanged and tasks such as \ac{IC} are performed \cite{sit2018}.

In this article, a design of binary preambles for synchronization is proposed and a pilot arrangement for channel, Doppler shift, and residual \ac{SFO} estimation is discussed for the considered \ac{PMCW}-based \ac{RadCom} system. Proof-of-concept measurement results are finally presented to validate the proposed synchronization strategy.

%

\section{System Model}\label{sec:sysModel}

Let a \ac{SISO} \ac{PMCW}-based radar system be based on the transmission of a \ac{PRBS} of length \mbox{$N\in\mathbb{N}_{>0}$}. More specifically, an m-sequence, also known as \ac{MLS}, is used, which results in \mbox{$N=2^m-1|m\in\mathbb{N}_{\geq2}$}. Every set of \mbox{$A\in\mathbb{N}_{>0}$} identical \acp{PRBS} constitutes a block, which is evaluated at the receiver side to yield a range profile. This is achieved by (i) discarding the first \ac{PRBS} within the block at the receiver side, letting it act as a \ac{CP}, (i) accumulating the remaining $A-1$ \acp{PRBS} repetitions, and (iii) performing a circular correlation between the resulting $N$-length accumulated \ac{PRBS} and the originally transmitted reference \ac{PRBS}. With the transmission of a frame containing \mbox{$M\in\mathbb{N}_{>0}$} blocks, \acp{DFT} can be performed along the bins of the range profiles to estimate Doppler shifts and ultimately generate a range-velocity radar image.

To enable \ac{RadCom} operation of such \ac{PMCW}-based radar system while still transmitting only binary sequences, each of the $M$ blocks in the \ac{PMCW} frame is modulated with a single \ac{BPSK} symbol \cite{giroto2021_tmtt,giroto_2021_PMCW}. At the receiver side, of another \ac{PMCW}-based \ac{RadCom} system, the same processing encompassing accumulation and circular correlation with the reference \ac{PRBS} as in the radar case is performed. After equalization, this results in $M$ blocks containing the autocorrelation of the reference \ac{PRBS} modulated by the corresponding \ac{BPSK} symbols, which can be extracted from the autocorrelation peak since it experiences the most processing gain, i.e., $N(A-1)$ \cite{giroto2021_tmtt}.

At the transmitter side of the considered \ac{PMCW}-based \ac{RadCom} system, the $NAM$ \acp{PRBS} in the transmit frame are generated by a single \ac{LFSR} with sampling rate $F_\mathrm{s}$ and then \ac{BPSK}-modulated. The discrete-time domain sequence output by the modulator is denoted by $x[n]\in\mathbb{C}$, $n\in\mathbb{Z}$ and has an equivalent continuous-time domain baseband transmit signal $x(t)\in\mathbb{C}$. Besides undergoing analog conditioning, e. g., to filter out \ac{OOB} emission, $x(t)\in\mathbb{C}$ is upconverted to the carrier frequency $f_\mathrm{c}\gg B$ and radiated by the transmit antenna. The aforementioned signal propagates through the medium and is eventually received by the receive antenna of a second \ac{PMCW}-based \ac{RadCom} system. The receive signal is then conditioned and downconverted to the baseband in an I/Q \ac{AFE}, resulting in the continuous-time domain signal without noise $\tilde{y}(t)\in\mathbb{C}$. It is henceforth assumed that $\tilde{y}(t)$ is the result of the propagation of $x(t)$ through a stronger, main path and multiple secondary paths, each with own delay, Doppler shift and phase rotation. Additionally, $\tilde{y}(t)$ contains the effects of mismatches between the transmitting and the receiving \ac{PMCW}-based \ac{RadCom} systems. Those are \ac{STO} due to distinct time references, besides \ac{CFO} and its resulting \ac{CPO} raised by the use of distinct oscillators by the non-collocated transmitter and receiver pair. To generate a discrete-time domain receive sequence $y[n]\in\mathbb{C}$ at the receiver side, the noise-impaired version of $\tilde{y}(t)$ undergoes \ac{A/D} conversion with a sampling rate set to the same value as at the transmitter, i.e., $F_\mathrm{s}$. However, as the sampling clock at the receiver is asynchronous w.r.t. the one at the transmitter, \ac{SFO} occurs.

To extract \ac{BPSK} data from $y[n]$, one must perform (i) synchronization to compensate for \ac{CFO}, \ac{STO}, and \ac{SFO}, (ii) compensation of Doppler shifts, and (iii) channel estimation and equalization to compensate for multipath propagation. The aforementioned task (i) can be performed via processing based on the transmission of a known preamble. After synchronization, a receive \ac{PMCW} frame can be formed and the last $A-1$ \acp{PRBS} of each block accumulated. The resulting blocks can be then circularly correlated with the reference \ac{PRBS}, and channel response and Doppler shift estimates can be obtained from the pilot blocks. Finally, tasks (ii) and (iii) are performed, i.e., both Doppler shift and \ac{CFR} are compensated, and the receive \ac{BPSK} symbols are extracted from the non-pilot or \ac{PL} blocks.

\section{Preamble Design and Pilot Arrangement}\label{sec:preambPilot}

In this article, a preamble is designed to meet requirements of both the \ac{SC} \cite{schmidl1997} (for time and frequency synchronization) and the Tsai \cite{tsai2005} (for sampling frequency synchronization) algorithms typically used in \ac{OFDM}-based systems, while equally-spaced, reserved blocks modulated with known \ac{BPSK} symbols at the receiver are used as pilots. To keep the transmitter architecture simple and linearity requirements low, both the preamble and pilots are required to be binary. The design of a binary preamble that meets the requirements of the adopted synchronization algorithms and the arrangement of pilots are discussed in Sections~\ref{subsec:preamble} and  \ref{subsec:pilot}, respectively.

\subsection{Binary Preamble Design}\label{subsec:preamble}

To perform time and frequency synchronization in the considered \ac{PMCW}-based \ac{RadCom} system, compensating both \ac{STO} and \ac{CFO}, $M_\mathrm{S\&C}=2$ binary preamble blocks that meet the requirements of the \ac{SC} algorithm are used \cite{schmidl1997}. Additionally, a sequence of $M_\mathrm{SFO}\in\mathbb{N}_{>0}$ \acp{PRBS} is designed to meet the requirements of the Tsai algorithm for \ac{SFO} estimation. Their design requirements is discussed as follows.

\subsubsection{Preamble Blocks for Schmidl \& Cox Algorithm}\label{subsubsec:preambleSC}

For the \ac{SC} algorithm, a preamble containing $M_\mathrm{S\&C}=2$ blocks of length $N^\mathrm{S\&C}_\mathrm{block}\in\mathbb{N}_{>0}$ is necessary. Disregarding \ac{CP}, one typically adopts $N^\mathrm{S\&C}_\mathrm{block}=N$ in \ac{OFDM} systems and designs the first block so that it consists of two equal halves, which results in a discrete-frequency domain spectrum with interleaved active tones. The second block must be designed so that at least the same tones as in the first preamble block are active, although usually all tones are active for \ac{OFDM}.

\begin{figure}[!t]
	\centering
	
	\resizebox{8cm}{!}{
		
		\psfrag{A}[c][c]{\footnotesize $N_\mathrm{S\&C}$}
		\psfrag{B}[c][c]{\footnotesize $N_\mathrm{SFO}$}
		
		\psfrag{C}[c][c]{\small Main path delay + STO}
		
		\psfrag{n}[c][c]{\footnotesize $n$}
		\psfrag{M[n]}[c][c]{\footnotesize $\gamma[n]$}
		\psfrag{r[n]}[c][c]{\scriptsize \textcolor{colorSC}{$r[n]$}}
		\psfrag{r[n+N]}[c][c]{\scriptsize \textcolor{colorSC}{$r[n+N_\mathrm{S\&C}]$}}
		\psfrag{0}[c][c]{\footnotesize $0$}
		
		\includegraphics[width=8.5cm]{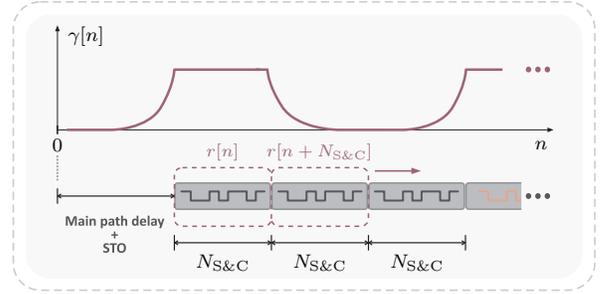}
	}
	\captionsetup{justification=raggedright,labelsep=period,singlelinecheck=false}
	\caption{\ Sliding window correlation and \ac{PRBS}-based \ac{SC} preamble.}\label{fig:schmidlCox}
	\vspace{-0.275cm}
\end{figure}

It is henceforth assumed that the length $N^\mathrm{S\&C}_\mathrm{block}$ of the two preamble blocks used for the \ac{SC} algorithm is $N^\mathrm{S\&C}_\mathrm{block}=3N_\mathrm{S\&C}$, where $N_\mathrm{S\&C}\in\mathbb{N}_{>0}$ is the length of a \ac{PRBS} such that \mbox{$N_\mathrm{S\&C}=[(N+1)/2]-1$}. In other words, each preamble block is composed of three repetitions of an \ac{MLS} whose length is approximately the half of the \ac{PL} \ac{PRBS} length $N$. The first repetition acts as a \ac{CP}, while the latter two are the required equal halves for the \ac{SC} algorithm. Although these equal halves are only indispensable for the first \ac{SC} block, the same structure is kept for the second one to keep the same block length as the \ac{PRBS} lengths are not flexible. The use of the considered preamble allows to find the start point of the \ac{PMCW} frame at the receiver via a sliding-window correlation. As illustrated in Fig.~\ref{fig:schmidlCox}, a correlation-like processing is performed between two sliding windows of length $N_\mathrm{S\&C}$ starting at the reference sample $n$ from $y[n]$, namely $r[n]$ and $r[n+N_\mathrm{S\&C}]$. The result is expressed as
\begin{equation}
\gamma[n] = \frac{\left|\sum_{\eta=0}^{N_\mathrm{S\&C}-1}r^*[n+\eta]r[n+\eta+N_\mathrm{S\&C}]\right|^2}{\sum_{\eta=0}^{N_\mathrm{S\&C}-1}\left|r[n+\eta+N_\mathrm{S\&C}]\right|^2}.
\end{equation}
Assuming orthogonal \acp{PRBS} for each \ac{SC} preamble block, which requires distinct \acp{LFSR}, $\gamma[n]$ will present two spaced plateaus of length $N_\mathrm{S\&C}$, each associated with one of the \ac{SC} preamble blocks. Considering only the first plateau, the frame start point can be estimated as one of the sample positions within the plateau \cite{schmidl1997}. Additionally, the phase of the term \mbox{$\sum_{\eta=0}^{N_\mathrm{S\&C}-1}r^*[n+\eta]r[n+\eta+N_\mathrm{S\&C}]$} in the numerator of $\gamma[n]$ can be evaluated to estimate the fractional \ac{CFO} (non-integer multiple of $F_\mathrm{s}/(2N_\mathrm{S\&C})$), and the first and second preamble blocks can be compared to estimate the integer \ac{CFO} (integer multiple of $F_\mathrm{s}/(2N_\mathrm{S\&C})$), also as discussed in \cite{schmidl1997}.

\subsubsection{Preamble Blocks for Tsai Algorithm}\label{subsubsec:preambleSFO}

In \cite{tsai2005}, multiple pairs of preamble \ac{OFDM} symbols are used to estimate the \ac{SFO} via weighted least-squares estimation and enable its correction with a resampling algorithm. In all these pairs, the second \ac{OFDM} symbol in the discrete-frequency domain is equal to the multiplication of first by a known \ac{PRBS}. In the considered \ac{PMCW}-based \ac{RadCom} system, however, changing the spectral content of a \ac{LFSR}-generated \ac{PRBS} is not possible. Instead, pairs of identical \acp{PRBS} are used. To avoid the need for extra shift registers, these are chosen to be identical to the one that is later modulated and repeated in the \ac{PL}. The use of identical pairs of identical \acp{PRBS} implies in the fact that no \ac{CP} is needed between the aforementioned \ac{PRBS} pairs. Therefore, a single copy of the adopted \ac{PRBS} is prepended to the following \acp{PRBS} pairs to act as \ac{CP} and avoid interblock interference from the previously transmitted \ac{SC} preamble blocks, which results in an odd total number of used \acp{PRBS} used for \ac{SFO} estimation $M_\mathrm{SFO}$.

Based on the proposed preamble design, one can conclude that a total of three \acp{LFSR} are needed in the considered \ac{PMCW}-based \ac{RadCom} system: two to generate $N_\mathrm{S\&C}$-length \acp{PRBS} for the \ac{SC} algorithm, and one to generate the $N$-length \ac{PRBS} used for both the Tsai algorithm and the \ac{PL}.

\subsection{Block Pilot Arrangement}\label{subsec:pilot}

\begin{figure}[!t]
	\centering
	\resizebox{8cm}{!}{
		\psfrag{A}[c][c]{\footnotesize $0$}
		\psfrag{B}[c][c]{\footnotesize $1$}
		\psfrag{H}[c][c]{}
		\psfrag{I}[c][c]{}
		\psfrag{J}[c][c]{}
		\psfrag{K}[c][c]{\footnotesize $NA-1$}
		
		\psfrag{EE}[c][c]{\footnotesize $0$}
		\psfrag{FF}[c][c]{\footnotesize $1$}
		\psfrag{GG}[c][c]{\footnotesize $2$}
		\psfrag{M-3}[c][c]{\footnotesize $M-1-\Delta M_\mathrm{pil}$}
		\psfrag{M-2}[c][c]{\footnotesize $M-\Delta M_\mathrm{pil}$}
		\psfrag{M-1}[c][c]{\footnotesize $M-\Delta M_\mathrm{pil}+1$}
		\psfrag{M-0}[c][c]{\footnotesize $M-1$}
		
		\psfrag{WWW}[c][c]{}
		\psfrag{ZZZ}[c][c]{\footnotesize $\Delta M_\mathrm{pil}$}
		
		\psfrag{x}[c][c]{}
		\psfrag{y}[c][c]{}
		
		\includegraphics[width=8.5cm]{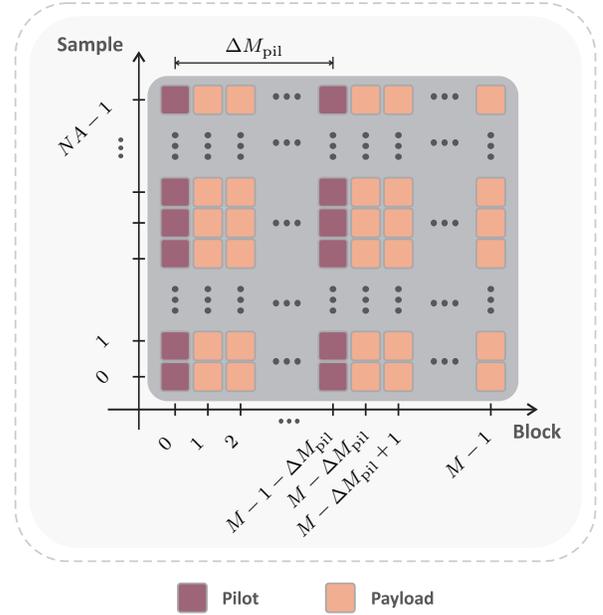}
	}
	\captionsetup{justification=raggedright,labelsep=period,singlelinecheck=false}
	\caption{\ Structure of the transmit PMCW frame containing \ac{PL} and pilots in the discrete-time domain. For simplicity, the preamble is ommitted.}\label{fig:frameStructure}
	\vspace{-0.275cm}
\end{figure}

A set of \ac{PL} blocks spaced by $\Delta M_\mathrm{pil}\in\mathbb{N}_{>0}$ blocks in the transmit \ac{PMCW} frame is reserved as depicted in Fig.~\ref{fig:frameStructure} and converted into pilot blocks by setting the \ac{BPSK} symbol that modulates than to $1$. Assuming that the pilot arrangement is known at the receiver, these blocks are accumulated and further processed to yield channel estimates for equalization and estimation and correction of the Doppler shift at the main path. The use of entire blocks with a block spacing of $\Delta M_\mathrm{pil}$ enables estimating channel responses with a maximum delay
\begin{equation}\label{eq:tau_max}
\tau_\mathrm{max}=N/F_\mathrm{s},
\end{equation}
and a maximum absolute Doppler shift at the main path \cite{sit2018}
\begin{equation}\label{eq:fD_max}
f_\mathrm{D,max}=\left[2(NA/F_\mathrm{s})\Delta M_\mathrm{pil}\right]^{-1}.
\end{equation}
Since even slight biases of the \ac{SFO} estimate with the Tsai algorithm result in performance degradation due to the accumulation of the \ac{SFO} effect over a high number of blocks, the obtained \ac{CIR} estimates with pilots can also be used to calculate and correct the residual \ac{SFO}. This is achieved by compensating the linearly increasing delay along the blocks caused by the residual \ac{SFO} \cite{burmeister2021}.

\section{Measurement Setup and Results}\label{sec:measResults}

In this section, proof-of-concept measurements are presented to validate the proposed \ac{PMCW}-based \ac{RadCom} system. In this context, a setup consisting of two Zynq UltraScale+ RFSoC ZCU111 \ac{SoC} platforms from Xilinx, Inc, was adopted. While the first ZCU111 emulated the transmitter of a \ac{PMCW}-based \ac{RadCom} system, the second ZCU 111 acted as the receiver for another \ac{PMCW}-based \ac{RadCom} system, both with $F_\mathrm{s}=\SI{1}{\giga\hertz}$. The \acp{DAC} of the transmitter board were directly connected to the \acp{ADC} of the receiver board via coaxial cables. Since the transfer functions of \acp{DAC}, \acp{ADC}, and cables were not calibrated, a channel with a main, stronger \ac{LoS} path and more severely attenuated secondary paths between transmitter and receiver was emulated at an \ac{IF} of $\SI{1}{\giga\hertz}$. Despite the fact that \ac{RF} \acp{AFE} were not used, \ac{STO}, \ac{CFO}, and \ac{SFO} could be experienced as the transmitter and receiver boards had distinct time, frequency, and sampling clock references.

A total of four distinct parameterizations, henceforth referred to as \ac{PMCW} \#1 to \#4, were adopted for the \ac{PMCW}-based \ac{RadCom} system aiming \ac{HAD} applications. All of them were defined seeking to achieve roughly the same dwell time of approximately $\SI{10.50}{\milli\second}$ for the \ac{PMCW} frame including preamble. The considered \ac{RF} parameters, \ac{PMCW} signal parameters, and the resulting communication and radar performance parameters, the latter calculated based on \cite{giroto2021_tmtt,giroto_2021_PMCW}, are listed in Table~\ref{tab:resultsParameters}. For conciseness, parameters that can be derived from already listed ones have been omitted. Additionally, the emulated carrier frequency of $\SI{79}{\giga\hertz}$ was only used to define the Doppler shifts that would result in the calculated radar velocity resolution and unambiguity. Since no additional Doppler shift to the \ac{CFO} was added in the measurements, the pilots were rather used to estimate a residual \ac{CFO} after correction with the \ac{SC} algorithm. The results from Table~\ref{tab:resultsParameters} show that, keeping $A$ and increasing $N$ yields higher communication processing gain $G_\mathrm{p,comm}=(N(A-1))$ and maximum delay for communication, while both the maximum Doppler shift and the data rate decrease due to the longer block duration. Similarly, radar maximum unambiguous range and velocity increase and decrease, respectively, along with $N$.

\begin{table}[!t]
	\renewcommand{\arraystretch}{1.5}
	\arrayrulecolor[HTML]{708090}
	\setlength{\arrayrulewidth}{.1mm}
	\setlength{\tabcolsep}{4pt}
	
	\centering
	\captionsetup{justification=centering}
	\caption{PMCW-based RadCom system parameters.}
	\label{tab:resultsParameters}
	\resizebox{\columnwidth}{!}{
		\begin{tabular}{|cc|c|c|c|}
			\hhline{|=====|}
			\multicolumn{1}{|c|}{} & \textbf{PMCW \#1} & \textbf{PMCW \#2} & \textbf{PMCW \#3} & \textbf{PMCW \#4} \\ \hhline{|=====|}
			\multicolumn{5}{|c|}{\textbf{PMCW signal parameters}} \\ \hhline{|=====|}
			\multicolumn{1}{|c|}{\textbf{Emulated carrier freq.} ($f_\mathrm{c}$)}  & \multicolumn{4}{c|}{$\SI{79}{\giga\hertz}$} \\ \hline
			\multicolumn{1}{|c|}{\textbf{Sampling frequency} ($F_\mathrm{s}$)}  & \multicolumn{4}{c|}{$\SI{1}{\giga\hertz}$} \\ \hline
			\multicolumn{1}{|c|}{\textbf{PRBS length} ($N$)}  & $255$ & $511$ & $1023$ & $2047$ \\ \hline
			\multicolumn{1}{|c|}{\textbf{PRBS repetitions for acc.} ($A$)}      & \multicolumn{4}{c|}{$5$} \\ \hline
			\multicolumn{1}{|c|}{\textbf{PL and pilot blocks} ($M$)}      & $8192$ & $4096$ & $2048$ & $1024$ \\ \hline
			\multicolumn{1}{|c|}{\textbf{\ac{SC} PRBS length} ($N_\mathrm{S\&C}$)} & $127$ & $255$ & $511$ & $1023$ \\ \hline
			\multicolumn{1}{|c|}{\textbf{PRBS for SFO est.} ($M_\mathrm{SFO}$)}      & \multicolumn{4}{c|}{$21$} \\ \hline
			\multicolumn{1}{|c|}{\textbf{Pilot block spacing} ($\Delta M_\mathrm{pil}$)}      & \multicolumn{4}{c|}{$5$} \\ \hhline{|=====|}
			\multicolumn{5}{|c|}{\textbf{Communication performance parameters}} \\ \hhline{|=====|}
			\multicolumn{1}{|c|}{\textbf{Comm. process. gain} ($G_\mathrm{p,comm}$)}  & $\SI{30.09}{dB}$  & $\SI{33.10}{dB}$ & $\SI{36.12}{dB}$ & $\SI{39.13}{dB}$ \\ \hline
			\multicolumn{1}{|c|}{\textbf{Maximum delay} ($\tau_\mathrm{max}$)} & $\SI{0.26}{\micro\second}$ & $\SI{0.51}{\micro\second}$ & $\SI{1.02}{\micro\second}$ & $\SI{2.05}{\micro\second}$ \\ \hline
			\multicolumn{1}{|c|}{\textbf{Max. Doppler shift} ($f_\mathrm{D,max}$)} & $\SI{78.43}{\kilo\hertz}$ & $\SI{39.14}{\kilo\hertz}$ & $\SI{19.55}{\kilo\hertz}$ & $\SI{9.77}{\kilo\hertz}$ \\ \hline
			\multicolumn{1}{|c|}{\textbf{Data rate} (100\% duty cycle)}      & $\SI{627.03}{kbit/s}$ & $\SI{312.67}{kbit/s}$ & $\SI{156.00}{kbit/s}$ & $\SI{77.78}{kbit/s}$ \\ \hhline{|=====|}
			\multicolumn{5}{|c|}{\textbf{Radar performance parameters}} \\ \hhline{|=====|}
			\multicolumn{1}{|c|}{\textbf{Radar process. gain} ($G_\mathrm{p,rad}$)}  & $\SI{69.22}{dB}$  & $\SI{69.23}{dB}$ & $\SI{69.23}{dB}$ & $\SI{69.23}{dB}$ \\ \hline
			\multicolumn{1}{|c|}{\textbf{Range resolution} ($\Delta R$)}     & \multicolumn{4}{c|}{$\SI{0.15}{\meter}$} \\ \hline
			\multicolumn{1}{|c|}{\textbf{Max. unamb. range} ($R_\mathrm{max,ua}$)}  & $\SI{38.22}{\meter}$  & $\SI{76.60}{\meter}$ & $\SI{153.34}{\meter}$ & $\SI{306.84}{\meter}$\\ \hline
			\multicolumn{1}{|c|}{\textbf{Velocity resolution} ($\Delta v$)}  & \multicolumn{4}{c|}{$\SI{0.18}{\meter/\second}$} \\ \hline
			\multicolumn{1}{|c|}{\textbf{Max. unamb. velocity} ($v_\mathrm{max,ua}$)}  & $\SI{744.09}{\meter/\second}$  & $\SI{371.32}{\meter/\second}$ & $\SI{185.48}{\meter/\second}$ & $\SI{92.69}{\meter/\second}$ \\ \hline
			\hhline{|=====|}
		\end{tabular}
	}
	\vspace{-0.275cm}
\end{table}

\begin{figure}[!t]
	\centering
	
	\psfrag{55}{(a)}
	\psfrag{22}{(b)}
	\psfrag{33}{(c)}
	\psfrag{44}{(d)}
	
	\psfrag{-2}[c][c]{\small -$2$}
	\psfrag{-1}[c][c]{\small -$1$}
	\psfrag{0}[c][c]{\small $0$}
	\psfrag{1}[c][c]{\small $1$}
	\psfrag{2}[c][c]{\small $2$}
	
	\psfrag{I}{$I$}
	\psfrag{Q}{$Q$}
	
	\psfrag{MER = EEEE}[c][c]{\scalebox{.9}{\scriptsize $\mathrm{MER}=\SI[parse-numbers = false]{28.53}{dB}$}}
	\psfrag{MER = BBBB}[c][c]{\scalebox{.9}{\scriptsize $\mathrm{MER}=\SI[parse-numbers = false]{34.48}{dB}$}}
	\psfrag{MER = CCCC}[c][c]{\scalebox{.9}{\scriptsize $\mathrm{MER}=\SI[parse-numbers = false]{33.95}{dB}$}}
	\psfrag{MER = DDDD}[c][c]{\scalebox{.9}{\scriptsize $\mathrm{MER}=\SI[parse-numbers = false]{-3.42}{dB}$}}
	
	\psfrag{BER = EEEE}[c][c]{\scalebox{.9}{\scriptsize $\mathrm{BER}=0$}}
	\psfrag{BER = BBBB}[c][c]{\scalebox{.9}{\scriptsize $\mathrm{BER}=0$}}
	\psfrag{BER = CCCC}[c][c]{\scalebox{.9}{\scriptsize $\mathrm{BER}=0$}}
	\psfrag{BER = DDDD}[c][c]{\scalebox{.9}{\scriptsize $\mathrm{BER}=0.50$}}
	
	\includegraphics[width=7cm]{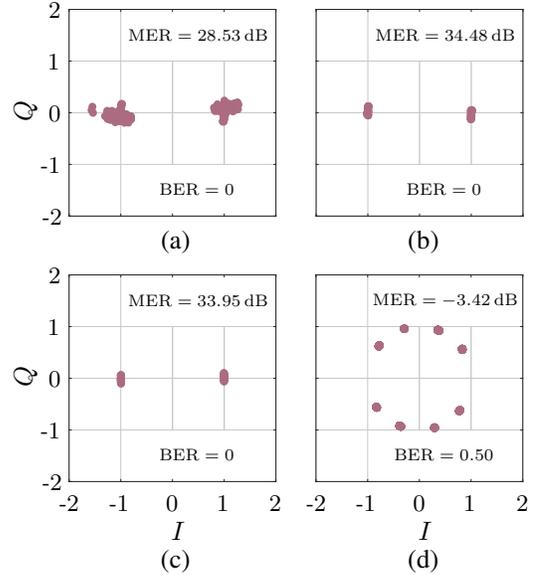}
	\captionsetup{justification=raggedright,labelsep=period,singlelinecheck=false}
	\caption{\ Superimposed normalized receive BPSK constellations from all measurement realizations for (a) PMCW \#1, (b) \#2, (c) \#3, and (d) \#4.}\label{fig:const_all}
	\vspace{-0.275cm}
\end{figure}

Finally, multiple measurements for PMCW \#1 to \#4 were performed with an \ac{SNR} of around $\SI{16.23}{dB}$. The achieved mean and standard deviation of the estimated synchronization mismatches in Table~\ref{tab:resultsSync} show that the precision of \ac{CFO} and \ac{SFO} estimates with the \ac{SC} and Tsai algorithms, respectively, increases with $N$ and $N_\mathrm{S\&C}$, i.e., from PMCW \#1 to \#4, which agrees with \cite{schmidl1997} and \cite{tsai2005}. The changes in the mean values are explained by the drift of time, frequency and sampling references during the measurements. Regarding residual \ac{CFO} estimates with pilots, PMCW \#1 to \#3 achieved sufficient accuracy, while PMCW \#4 estimated a strongly biased residual \ac{CFO}, i.e., $\SI{5.86}{\kilo\hertz}$ instead of around $\SI{-15}{\kilo\hertz}$, due to its constraining to $\SI{\pm9.77}{\kilo\hertz}$. Next, Fig.~\ref{fig:const_all} shows the superimposed normalized receive \ac{BPSK} constellations from the all measurements and their corresponding \ac{MER} and \ac{BER} values. Although the obtained \acp{MER} are not equal to the input \ac{SNR} plus the expected $G_\mathrm{p,comm}$, which is in part due to imperfections in synchronization and channel estimation, a tendency of increasing \ac{MER} along with $G_\mathrm{p,comm}$ is observed from PMCW \#1 to \#2. Between PMCW \#2 and \#3, the \ac{MER} slightly decreases mainly due to the accumulated effect of residual \ac{CFO} and \ac{SFO} between largely spaced pilots even after dual compensation, which degrades the quality of the obtained channel matrix used for equalization. Finally, a negative \ac{MER} was achieved by PMCW \#4 due to its ambiguous residual \ac{CFO} estimate, which led to rotations of the obtained constellations w.r.t. the expected ones and resulted in a \ac{BER} of $0.50$.
%


\section{Conclusion}\label{sec:conclusion}

This article introduced a strategy for synchronization and channel estimation in radar-centric \ac{PMCW}-based \ac{RadCom} systems based on \acp{PRBS} generated by \acp{LFSR}. In this context, the design of compatible preambles with the \ac{SC} and Tsai algorithms typically used for time, frequency and sampling frequency synchronization in \ac{OFDM}, and the arrangement of pilots for channel, Doppler shift, and residual \ac{SFO} estimation were discussed. Finally, proof-of-concept measurements validated the proposed strategy, showing that a \ac{PMCW}-based \ac{RadCom} system achieves robust communication performance if correctly parameterized and same sensing performance to an equally-parameterized \ac{PMCW}-based radar system.

\begin{table}[!t]
	\renewcommand{\arraystretch}{1.5}
	\arrayrulecolor[HTML]{708090}
	\setlength{\arrayrulewidth}{.1mm}
	\setlength{\tabcolsep}{4pt}
	
	\centering
	\captionsetup{justification=centering}
	\caption{Synchronization performance results. Mean value and standard deviation of estimated parameters over multiple measurements are shown.}
	\label{tab:resultsSync}
	\resizebox{\columnwidth}{!}{
		\begin{tabular}{|cc|c|c|c|}
			\hhline{|=====|}
			\multicolumn{1}{|c|}{} & \textbf{PMCW \#1} & \textbf{PMCW \#2} & \textbf{PMCW \#3} & \textbf{PMCW \#4} \\ \hhline{|=====|}
			\multicolumn{1}{|c|}{\textbf{\ac{CFO} via \ac{SC} ($\SI{}{\kilo\hertz}$)}}  & $-83.76\pm7.84$ & $-84.93\pm6.11$ & $-85.26\pm1.17$ & $-86.20\pm0.92$ \\ \hline
			\multicolumn{1}{|c|}{\textbf{Resid. \ac{CFO} via pilots ($\SI{}{\kilo\hertz}$)}}  & $-15.62\pm8.14$ & $-15\pm6.31$ & $-15.01\pm1.31$ & $5.86\pm0.80$ \\ \hline
			\multicolumn{1}{|c|}{\textbf{\ac{SFO} via Tsai (ppm)}}  & $98.38\pm5.98$ & $99.62\pm1.76$ & $100.28\pm0.91$ & $99.63\pm0.19$ \\ \hline
			\hhline{|=====|}
		\end{tabular}
	}
\end{table}

\section*{Acknowledgment}

The authors acknowledge the financial support by the Federal Ministry of Education and Research of Germany in the projects ``ForMikro-REGGAE'' (grant number: 16ES1061). The work of Lucas Giroto de Oliveira was also financed by the German Academic Exchange Service (DAAD) - Funding program 57440921/Pers. Ref. No. 91555731.\\

\bibliographystyle{IEEEtran}
\bibliography{./OverviewPapers,./RadCom_Enablement,./B5G_6G,./Interference,./Automotive,./RadarNetworks,./ChirpSequence,./PMCW,./OFDM,./OCDM,./OFDM_Variations,./CS_OCDM_Variations,./CP_DSSS,./BandwidthEnlargement_DigitalRadars,./CompressedSensing_DigitalRadars,./RadarTargetSimulator,./Parameters,./HardwareImplementation,./FirstRadCom,./Interference_CS,./ResourceAllocation,./SFO,./Bistatic}

\end{document}